\documentclass[aps,pra,amsmath,amssymb, reprint, superscriptaddress,showpacs]{revtex4-1}
\usepackage{color, graphicx}     
\usepackage{dcolumn}     
\usepackage{bm}                  
\usepackage{amssymb}
\usepackage{latexsym}
\usepackage{amsfonts}
\usepackage{amsmath}
\usepackage{multirow}
\usepackage{color,soul}
\usepackage[colorlinks, linkcolor= blue, citecolor = blue, urlcolor=blue]{hyperref}
\usepackage[dvipsnames]{xcolor}
\usepackage{times}
\begin{document}


\makeatletter
\newcommand*\bigcdot{\mathpalette\bigcdot@{.3}}
\newcommand*\bigcdot@[2]{\mathbin{\vcenter{\hbox{\scalebox{#2}{$\m@th#1\bullet$}}}}}
\makeatother

\title
{Gain-loss-engineering: a new platform for extreme anisotropic thermal photon tunneling}

\newcommand*{\IUF}[0]{{Institut Universitaire de France, 1 rue Descartes, F-75231 Paris, France}}
\newcommand*{\HIT}[0]{{School of Energy Science and Engineering, Harbin Institute of Technology, Harbin 150001, China}}
\newcommand*{\LCC}[0]{{Laboratoire Charles Coulomb (L2C), UMR 5221 CNRS-Université de Montpellier, F-34095 Montpellier, France}}
\newcommand*{\LAT}[0]{{Key Laboratory of Aerospace Thermophysics, Ministry of Industry and Information Technology, Harbin 150001, China}}
\newcommand*{\HNU}[0]{{School of Physics and Electronics, Hunan University, Changsha 410082, China}}
\newcommand*{\FWL}[0]{{Fields \& Waves Lab, Department of Engineering, University of Sannio, Benevento, I-82100, Italy}}

\author{Cheng-Long Zhou}
\affiliation{\HIT}%
\affiliation{\LAT}%
\author{Yu-Chen Peng}
\affiliation{\HNU}%
\author{Yong Zhang}
\affiliation{\HIT}
\affiliation{\LAT}%
\author{Hong-Liang Yi}
\email{Corresponding author: yihongliang@hit.edu.cn}
\affiliation{\HIT}%
\affiliation{\LAT}%
\author{Mauro Antezza}
\email{Corresponding author: mauro.antezza@umontpellier.fr}
\affiliation{\IUF}
\affiliation{\LCC}%
\author{Vincenzo Galdi}
\email{Corresponding author: vgaldi@unisannio.it}
\affiliation{\FWL}

\begin{abstract}
We explore a novel approach to achieving anisotropic thermal photon tunneling, inspired by the concept of parity-time symmetry in quantum physics. Our method leverages the modulation of constitutive optical parameters, oscillating between loss and gain regimes. This modulation reveals a variety of distinct effects in thermal photon behavior and dispersion. Specifically, we identify complex tunneling modes through gain-loss engineering, which include thermal photonic defect states and Fermi-arc-like phenomena, which surpass those achievable through traditional polariton engineering. Our research also elucidates the laws governing the evolution of radiative energy in the presence of gain and loss interactions, and highlights the unexpected inefficacy of gain in enhancing thermal photon energy transport compared to systems characterized solely by loss. This study not only broadens our understanding of thermal photon tunneling but also establishes a versatile platform for manipulating photon energy transport, with potential applications in thermal management, heat science, and the development of advanced energy devices.
\end{abstract}
\maketitle

 \setlength{\parskip}{0pt}
\noindent\textbf{1.\,Introduction}

\setlength{\parskip}{5pt}
When two objects are positioned closer together than the characteristic wavelength of thermal radiation, an intriguing phenomenon may occur: evanescent waves between the objects can enable the tunneling of energetic photons across the vacuum gap, a phenomenon known as thermal photon tunneling\,\cite{zhang2007nano}. This effect facilitates an electromagnetic (EM) energy transfer that significantly exceeds the blackbody radiation limit, often by several orders of magnitude\,\cite{carminati1999near, volokitin2007near, Song20016NN, Salihoglu2023PRL}. This super-Planckian property renders thermal photon tunneling an exciting prospect for groundbreaking advancements in energy conversion and thermal modulation devices, 
representing a significant breakthrough in the domains of thermal science and thermoengineering\,\cite{Zhu2019Nature, Ben2018NN, Oety2010PRL, ben2014near, kubytskyi2014radiative}.

\setlength{\parskip}{0pt}
Recent research has shown that the thermal photon tunneling effect displays a marked directional dependence in wavevector space, leading to what is known as anisotropic states of thermal photon tunneling. These states are particularly influenced by factors such as crystal structure, external fields, or confinement effects\,\cite{Liu2021Extrem, Tony2017NM}. These anisotropic states not only have the potential to further enhance radiative heat flux by facilitating EM energy transfer across larger wavevectors but also introduce a new dimension for precise and effective thermal manipulation\,\cite{tang2021PRL, Wang2024NC, Li2024NRM}. Traditionally, inducing anisotropic states in thermal photon tunneling has primarily relied on polariton engineering\,\cite{Fan2017PRL, Liu2021Extrem, Tony2017NM}. For instance, Liu {\em et al.} \cite{Liu2015ACS} achieved significant enhancements (more than tenfold) in heat flux by employing hyperbolic polariton engineering to introduce substantial anisotropy in thermal photon tunneling. Similarly, Zhou {\em et al.} \cite{Zhou2022PRB} made notable advances using shear polaritons to increase the anisotropy of thermal photon tunneling, thereby enhancing its heat manipulation capabilities.
The past decade has seen intense research focused on developing new structures and materials to advance polariton engineering, which has significantly deepened our understanding of anisotropic thermal photon tunneling and provided new methods for its modulation\,\cite{Zheng2022ACS,Du2023PRApp,Tang2021ACS,Luo2024arXiv,Luo2020PRB}. However, this raises an important question: Is polariton engineering the only approach to achieving anisotropic thermal photon tunneling, or are there alternative methods yet to be explored?
 
\setlength{\parskip}{0pt}
Recent advances, inspired by quantum-physics concepts such as "parity-time" (PT) symmetry, have sparked considerable interest in exploring systems characterized by gain and loss into various branches of physics
\cite{Konotop2016RMP, Ozdemir2019NM, Ramy2018NP}. In photonics, acoustics, and electronics, controlled modulation of gain and loss can drastically alter system responses, unlocking effects far beyond mere compensation. Research has highlighted the crucial role of the imaginary part of the refractive index, which parameterizes gain and loss, in modulating optical field strength\,\cite{Vashahri2019PRA} and achieving localized states \cite{Moccia2020ACS}, as well as extreme parameters\,\cite{Coppolaro2020PNAS} and anisotropy\,\cite{Coppolaro2020ACS,Galdi2021IEEE}.\,This dynamic has paved the way for a range of applications across scientific fields, including optical sensing\,\cite{Vashahri2019PRA}, laser technology\,\cite{Hui2014PRL}, amplifiers\,\cite{Koutserimpas2018PRL}, and signal processing\,\cite{Li2018PRL}. Furthermore, the effective manipulation of gain and loss as fundamental elements of non-Hermitian systems establishes a robust framework, enabling the exploration of diverse topological features in optics\,\cite{Hodaei2017Nature}, acoustics\,\cite{Helbig2021NP}, and electricity\,\cite{Zhu2018PRL}. This exploration into the intriguing capabilities of gain-loss modulation naturally leads us to contemplate whether it might be possible to harness this concept within the context of thermal photon tunneling.

\setlength{\parskip}{0pt}
\begin{figure*}
	\centering
	\centerline{\includegraphics[width=1\textwidth]{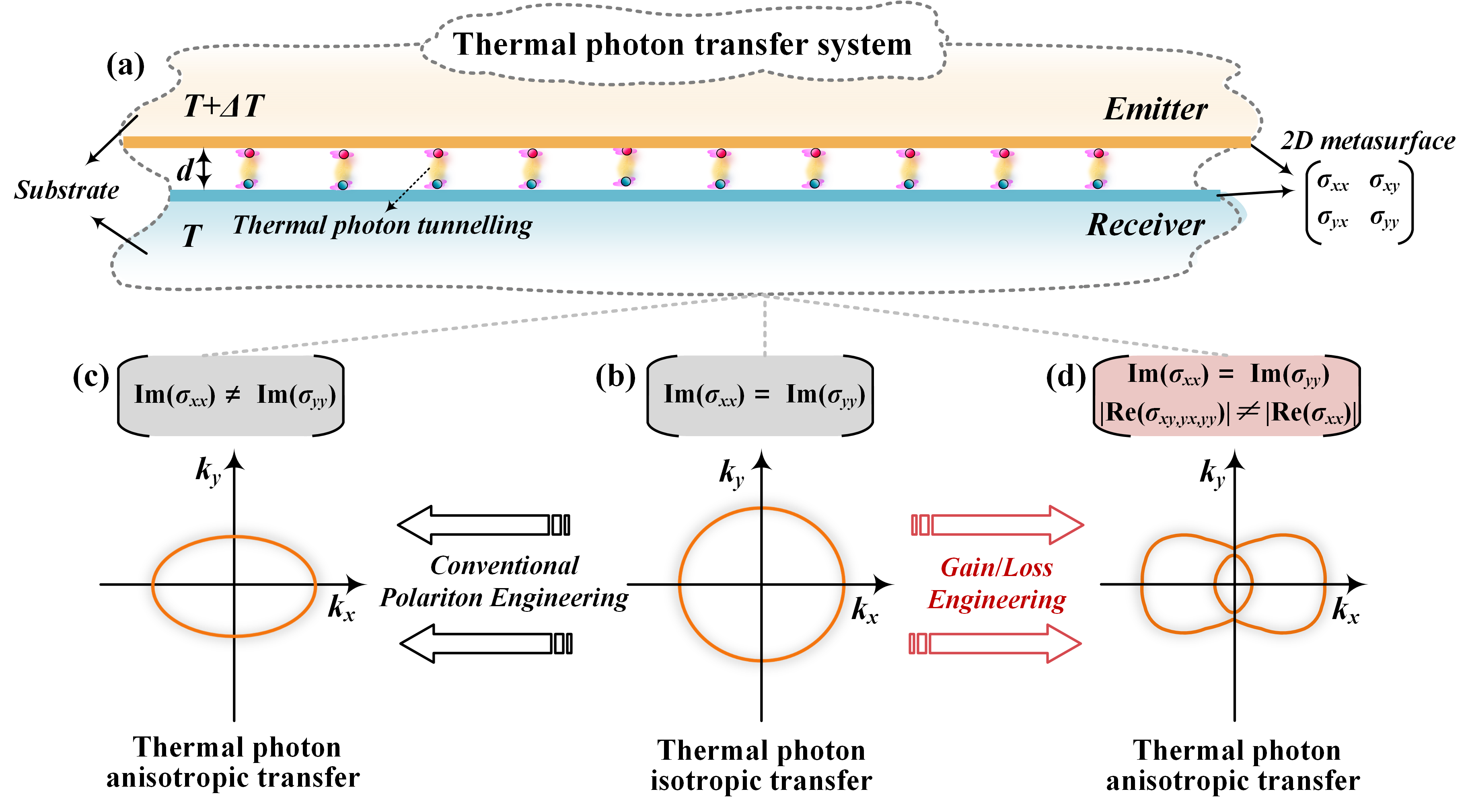}}
	\caption{Anisotropic thermal photon tunneling effect. (a) Schematic of thermal photon tunneling between metasurfaces exhibiting gain and loss separated by a vacuum gap of width $d$. (b), (c), (d) Representative isofrequency contours of thermal photon tunneling effect 
		for isotropic configuration
		with identical diagonal elements [$\mbox{Im}(\sigma_{xx})=\mbox{Im}(\sigma_{yy})$], conventionally anisotropic configuration with different diagonal elements [$\mbox{Im}(\sigma_{xx})\neq\mbox{Im}(\sigma_{yy})$],  and gain-loss engineering [$\mbox{Im}(\sigma_{xx})=\mbox{Im}(\sigma_{yy})$, $\mbox{Re}(\sigma_{xy,yx,yy}) \neq |\mbox{Re}(\sigma_{xx})|$], respectively.} 
	\label{Fig1}
\end{figure*}

\setlength{\parskip}{0pt}
Accordingly, we introduce a novel approach to achieve anisotropic thermal photon tunneling exclusively through the manipulation of gain and loss in metasurfaces. Our results demonstrate that judicious tailoring of the gain-loss interplay within the diagonal term of the optical conductivity tensor can induce anisotropic responses in a thermal photon tunneling system which vary significantly with the gain-loss level. Counterintuitively, and in stark contrast to conventional optical gain devices, the introduction of gain actually leads to a reduction in the energy transfer efficiency compared to purely lossy systems. This unexpected result is attributed to the formation of a thermal photon defect state (\textit{tp}-defect state). 
Further investigations into the evolutionary trajectory of thermal photon tunneling under gain-loss interactions within non-diagonal terms reveal the emergence of Fermi-arc-like and extremely anisotropic tunneling states. 

\setlength{\parskip}{0pt}
The rest of this paper is structured as follows. Section 2 outlines the problem formulation and its geometry. In Section 3, we identify and explore the parameter regimes of interest, and illustrate some representative results, obtained via fluctuational electrodynamics. Specifically, we study the anisotropic response of thermal photon tunneling under gain-loss interplay within the diagonal and non-diagonal terms of the optical conductivity tensor, and explore its counterintuitive law of energy transfer. Finally, Section 4 provides some brief conclusions and discusses potential future research directions.

\setlength{\parskip}{20pt}
\noindent\textbf{2.\,Problem Geometry and Formulation}
 
 \setlength{\parskip}{5pt}
\label{Sec2}
In our model, we examine thermal photon tunneling between two identical, parallel monolayer metasurfaces separated by a vacuum gap of width $d$ [see Fig.\,\ref{Fig1}(a)]. We begin by defining a metasurface characterized by loss and gain via an optical conductivity tensor
\setlength{\parskip}{0pt}
\begin{equation}
\label{eq:1}
\begin{aligned}
\sigma & =
\begin{bmatrix}
\sigma_{xx} & \sigma_{xy} \\
\sigma_{yx} & \sigma_{yy}
\end{bmatrix}
\\ & =
\begin{bmatrix}
\mathrm{Re}(\sigma_{D})+\textit{i}\mathrm{Im}(\sigma_{D}) & \eta_{xy}  \mathrm{Re}(\sigma_{D}) \\
\eta_{yx} \mathrm{Re}(\sigma_{D}) & \eta_{yy}  \mathrm{Re}(\sigma_{D})+\textit{i}\mathrm{Im}(\sigma_{D})
\end{bmatrix}.
\end{aligned}
\end{equation}
Here and henceforth, we assume an implicit $\exp(-i\omega t)$ time-harmonic dependence, so that positive and negative real parts in the surface conductivity indicate loss and gain, respectively \cite{Galdi2021IEEE}. 
In Eq.\,(\ref{eq:1}), loss and gain are parameterized through the coefficients $\eta_{xy,yx,yy}=\mathrm{Re}(\sigma_{xy,yx,yy})/\mathrm{Re}(\sigma_{xx})$, which control the real part of the individual components within the conductivity tensor \cite{Lannebere2022PRL}. 
We employ the conventional Drude model $\sigma_{D}=\sigma_{0} \omega_{f}/(\gamma-i\omega)$ as a foundational framework, assuming  $\sigma_0=e^{2}/\hbar$, $\omega_{f} = E_{f}/\hbar$, and $\gamma = 10^{12}$ rad/s, with $E_{f} = 0.2$ eV, and $e$ and $\hbar$ denoting the electron charge and reduced Plank's constant, respectively. This model serves as an excellent basis for thorough characterization and is extensively used to study complex EM mechanisms such as adjoint Kirchhoff’s law \cite{Guo2023PRX}, transistor-like optical gain \cite{Rappoport2023PRL}, and non-Hermitian responses \cite{Lannebere2022PRL}. In practical settings, effective conductivity models that incorporate gain and loss can be implemented through the artificial arrangement of material constituents featuring amplification or absorption \cite{Zheludev2008PRL, Xiao2010Nature, Kuzmin2021OL}.

It is important to note that gain and loss can coexist within the same system, and a composite conductivity that incorporates both can be achieved by integrating gain and loss elements oriented in different directions \cite{Kuzmin2021OL}. The gain elements can be activated using semiconductor quantum dots \cite{Kuzmin2021OL} and electrical response \cite{Rappoport2023PRL}. In this study, our focus is on analyzing the characteristics of EM energy transfer, specifically the photonic transmission coefficient. The emitter and receiver in our setup are maintained at temperatures \(T\) and \(T + \Delta T\), respectively, with \(T\) set at 300 K. This coefficient quantifies the probability of photon transfer/absorption driven by thermal energy, which can be expressed as follows \cite{Biehs2011APL}:
\setlength{\parskip}{0pt}
\begin{equation}
\label{eq:2}
\begin{aligned}
\xi=\begin{cases}
\mathrm{Tr}[(\textbf{I}-\textbf{R}^{\dagger}\textbf{R})\textbf{D}(\textbf{I}-\textbf{R}\textbf{R}^{\dagger})\textbf{D}^{\dagger}], & k\le k_{0},
\\
\mathrm{Tr}[(\textbf{R}^{\dagger}-\textbf{R})\textbf{D}(\textbf{R}-\textbf{R}^{\dagger})\textbf{D}^{\dagger}]e^{2ik_{z}d},      & k>k_{0}.
\end{cases}
\end{aligned}
\end{equation}

In Eq. (\ref{eq:2}), $\mbox{Tr}$ and $\dagger$ denote the trace and Hermitian adjoint, respectively. Moreover, $k_x$, $k_y$ and $k_z$ denote the $x$, $y$ and $z$ components of the wavevector, with $k=\sqrt{k^{2}_{x}+k^{2}_{y}}$ representing its in-plane wavevector magnitude and $k_{0}=\omega/c$ the wavenumber in vacuum; propagating and evanescent waves correspond to $\textit{k} \le \textit{k}_{0}$ and $\textit{k} > \textit{k}_{0}$, respectively. 
The matrix \(\textbf{D} = (\textbf{I} - \textbf{R}\textbf{R}e^{2ik_{z}d})^{-1}\) \cite{Luo2020APL} represents the conventional Fabry-Perot-like denominator arising from multiple scattering between the two interfaces of the emitter and receiver. Here, \(\textbf{I}\) denotes a \(2 \times 2\) identity matrix, and
\(\textbf{R}\) represents a \(2 \times 2\) reflection matrix applicable to all polarization representations. The dispersion of collective surface waves within such a system can be determined by setting the determinant of \(\textbf{D}\) equal to zero \cite{zhang2007nano}. In the homogenization approach, the EM response of such a metasurface can generally be characterized using a fully populated conductivity tensor in the wavevector space  \cite{Zhou2022PRApp},  viz.,
\begin{widetext}
\begin{align}
\tilde{\sigma}=
\begin{bmatrix}
\tilde{\sigma}_{xx} & \tilde{\sigma}_{xy} \\
\tilde{\sigma}_{yx} & \tilde{\sigma}_{yy}
\end{bmatrix}
=\frac{1}{k_{0}^{2}}
\begin{bmatrix}
k_{x}^{2}\sigma_{xx}+k_{y}^{2}\sigma_{yy}+k_{x}k_{y}(\sigma_{xx}+\sigma_{yy}) & k_{x}^{2}\sigma_{xy}-k_{y}^{2}\sigma_{yx}+k_{x}k_{y}(\sigma_{yy}-\sigma_{xx}) \\
k_{x}^{2}\sigma_{yx}-k_{y}^{2}\sigma_{xy}+k_{x}k_{y}(\sigma_{yy}-\sigma_{xx}) & k_{x}^{2}\sigma_{yy}+k_{y}^{2}\sigma_{xx}-k_{x}k_{y}(\sigma_{xy}-\sigma_{yx})
\end{bmatrix}.
\end{align}
\end{widetext} 

\setlength{\parskip}{0pt}
For \textit{p}-polarization, the EM field components are defined as \(E_{p} = [E_{x}, 0, E_{z}]\) and \(H_{p} = [0, H_{y}, 0]\); for \textit{s}-polarization, they are \(E_{s} = [0, E_{y}, 0]\) and \(H_{s} = [H_{x}, 0, H_{z}]\). By applying these \textit{p}- and \textit{s}-wave components to the boundary conditions at the metasurfaces, we can determine the relationships among all EM field components in the entire space  \cite{Kotov2019PRB}, viz.,
\begin{equation}
\label{eq:5}
\begin{bmatrix}
H_{pV}^{+} \\ H_{pV}^{-} \\
E_{sV}^{+} \\ E_{sV}^{-} 
\end{bmatrix}
=\textbf{T}
\begin{bmatrix}
H_{pS}^{+} \\ H_{pS}^{-} \\
E_{sS}^{+} \\ E_{sS}^{-} 
\end{bmatrix}
=
\begin{bmatrix}
T_{11} & T_{12} & T_{13} & T_{14}\\
T_{21} & T_{22} & T_{23} & T_{24}\\
T_{31} & T_{32} & T_{33} & T_{34}\\
T_{41} & T_{42} & T_{43} & T_{44}\\
\end{bmatrix}
\begin{bmatrix}
H_{pS}^{+} \\ H_{pS}^{-} \\
E_{sS}^{+} \\ E_{sS}^{-} 
\end{bmatrix}.
\end{equation}
In Eq. (\ref{eq:5}), the $+$ and $-$ signs represent forward (from vacuum gap to substrate) and backward waves, respectively, and the subscripts \textit{V} and \textit{S} indicate the media above and below the metasurface, respectively. The transfer matrix $\textbf{T}$ can be derived as \cite{Kotov2019PRB}
\begin{equation}       
\label{eq:6}
\textbf{T}
=\frac{1}{2}
\begin{pmatrix}
\frac{k_{zS}}{\varepsilon_{S}}
\begin{bmatrix}
P_{1} & P_{2}\\
P_{3} & P_{4}
\end{bmatrix}
&
\sqrt{\frac{\mu_{0}}{\varepsilon_{0}}}\tilde{\sigma}_{xy}
\begin{bmatrix}
1 & 1\\
1 & 1
\end{bmatrix}
\\
\sqrt{\frac{\mu_{0}}{\varepsilon_{0}}}\frac{k_{zS}\tilde{\sigma}_{yx}}{\varepsilon_{S}k_{zV}}
\begin{bmatrix}
1 & 1\\
1 & 1
\end{bmatrix}
&
\frac{1}{k_{zV}}
\begin{bmatrix}
S_{1} & S_{2}\\
S_{3} & S_{4}
\end{bmatrix}
\end{pmatrix},
\end{equation}
where the \textit{p}- and \textit{s}-waves components are \cite{Kotov2019PRB}
\setlength{\parskip}{0pt}
\begin{figure*}
	\centering
	\centerline{\includegraphics[width=1\textwidth]{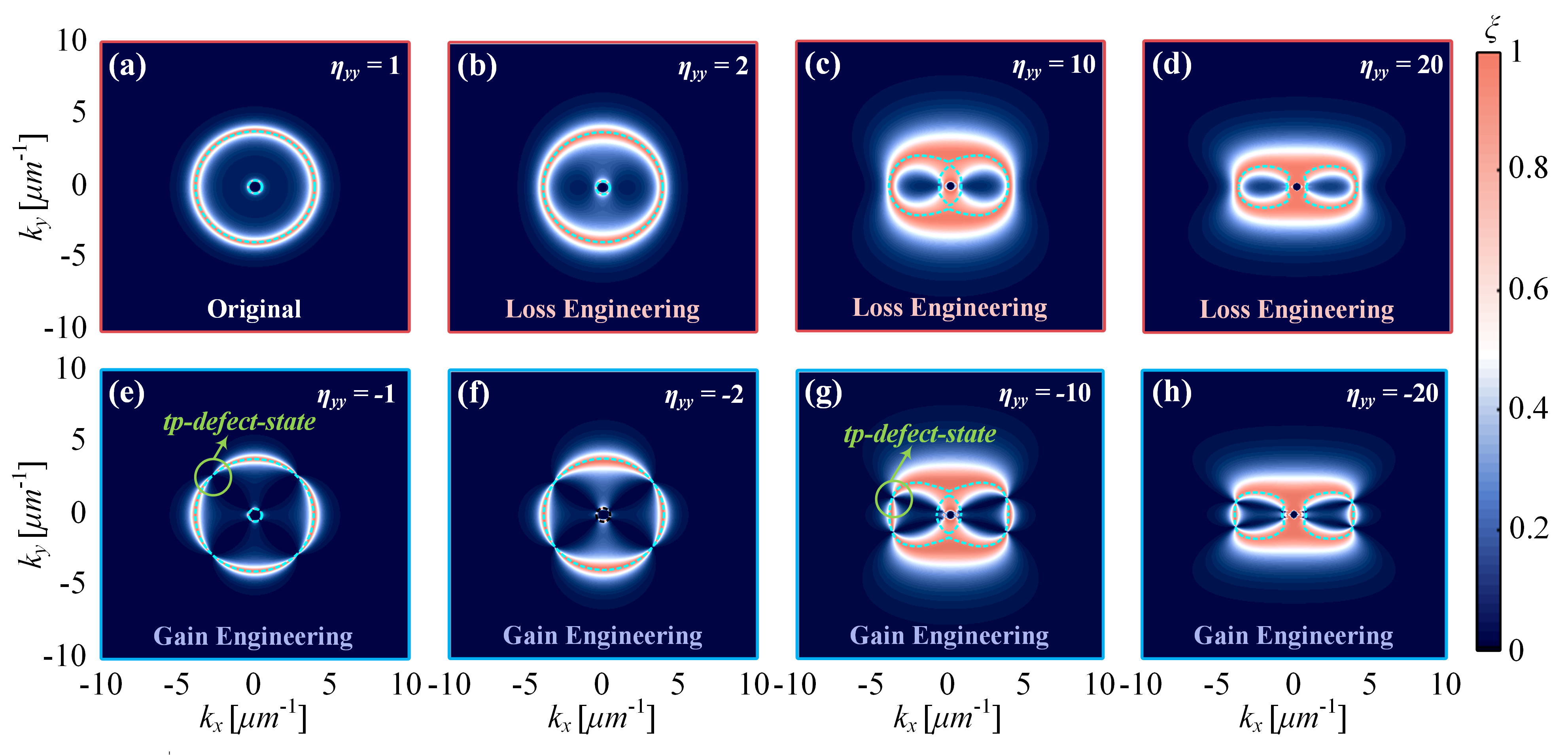}}
	\caption{Evolution of photonic transmission coefficient $\xi$ when applying gain and loss in the diagonal term of the conductivity tensor,
		parameterized through the coefficient $\eta_{yy}=\mbox{Re}(\sigma_{yy})/\mbox{Re}(\sigma_{xx})$.
		 (a) $\eta_{yy} = 1$ (original), (b) $\eta_{yy} = 2$, (c) $\eta_{yy} = 10$, (d) $\eta_{yy} = 20$, (e) $\eta_{yy} = -1$, (f) $\eta_{yy} = -2$, (g) $\eta_{yy} = -10$, and (h) $\eta_{yy} = -20$. The dashed curves in each panel represent the isofrequency contours.}
	\label{Fig2}
\end{figure*}
\setlength{\parskip}{0pt}
\begin{equation}
\label{eq:7}
\begin{aligned}
&P_{1}=\frac{1}{k_{zV}}+\frac{\varepsilon_{S}}{k_{zS}}+\frac{\tilde{\sigma}_{xx}}{\omega \varepsilon_{0}}, S_{1}=k_{zV}+k_{zS}+\omega \mu_{0} \tilde{\sigma}_{yy},\\ 
&P_{2}=\frac{\varepsilon_{S}}{k_{zs}}-\frac{1}{k_{zV}}-\frac{\tilde{\sigma}_{xx}}{\omega \varepsilon_{0}}, S_{2}=k_{zV}-k_{zS}+\omega \mu_{0} \tilde{\sigma}_{yy},\\
&P_{3}=\frac{\varepsilon_{S}}{k_{zs}}-\frac{1}{k_{zV}}+\frac{\tilde{\sigma}_{xx}}{\omega \varepsilon_{0}}, S_{3}=k_{zV}-k_{zS}-\omega \mu_{0} \tilde{\sigma}_{yy},\\
&P_{4}=\frac{1}{k_{zV}}+\frac{\varepsilon_{S}}{k_{zS}}-\frac{\tilde{\sigma}_{xx}}{\omega \varepsilon_{0}}, S_{4}=k_{zV}+k_{zS}-\omega \mu_{0} \tilde{\sigma}_{yy}.
\end{aligned}
\end{equation}
In Eqs. (\ref{eq:6}) and (\ref{eq:7}), $\varepsilon_{S}$ represents the relative permittivity of the substrate, which is commonly set to one, following conventional practice in the literature \cite{ilic2012near, ilic2018ACS, Yu2017NC}. The reflection matrix $\textbf{R}$ is defined below using the elements of the $\textbf{T}$-matrix \cite{Hu2020NL}:
\begin{equation}
\label{eq:8}
\textbf{R}=
\begin{bmatrix}
r_{pp} & r_{ps}\\
r_{sp} & r_{ss}
\end{bmatrix}
\end{equation}
with
\begin{equation}
\label{eq:9}
\begin{aligned}
&r_{pp}=\frac{T_{21}T_{33}-T_{23}T_{31}}{T_{11}T_{33}-T_{13}T_{31}},
r_{ps}=\frac{T_{41}T_{33}-T_{43}T_{31}}{T_{11}T_{33}-T_{13}T_{31}},\\
&r_{sp}=\frac{T_{11}T_{23}-T_{13}T_{21}}{T_{11}T_{33}-T_{13}T_{31}},
r_{ss}=\frac{T_{11}T_{43}-T_{13}T_{41}}{T_{11}T_{33}-T_{13}T_{31}}.
\end{aligned}
\end{equation}

\setlength{\parskip}{20pt}
\noindent\textbf{3.\,Results and Discussion}

\setlength{\parskip}{5pt}
\label{Sec3}
For $\mathrm{Im}(\sigma_{xx}) = \mathrm{Im}(\sigma_{yy})$, the system exhibits conventional isotropic circular dispersion, as illustrated in Fig.\,\ref{Fig1}(b) \cite{Alu2015PRL}. However, when polariton engineering is applied to the system, such that $\mathrm{Im}(\sigma_{xx}) \neq \mathrm{Im}(\sigma_{yy})$,
the system's dispersion takes on an anisotropic form, as shown in Fig.\,\ref{Fig1}(c).

\setlength{\parskip}{0pt}
In the presence of loss-gain-induced effects ($|\mathrm{Re}(\sigma_{xy,yx,yy})|\gg|\mathrm{Re}(\sigma_{xx})|$), even when the imaginary part of the diagonal element remains constant, the system can still exhibit anisotropic thermal photon tunneling phenomena, as illustrated in Fig.\,\ref{Fig1}(d). Importantly, these conditions lead to a richer and more complex variety of anisotropic states of thermal photon tunneling. This finding paves the way for innovative methods of manipulating thermal photons, introducing new levels of complexity and potential applications in the dynamics of thermal photons.

\begin{figure}[b]
	\centering
	\centerline{\includegraphics[width=1\columnwidth]{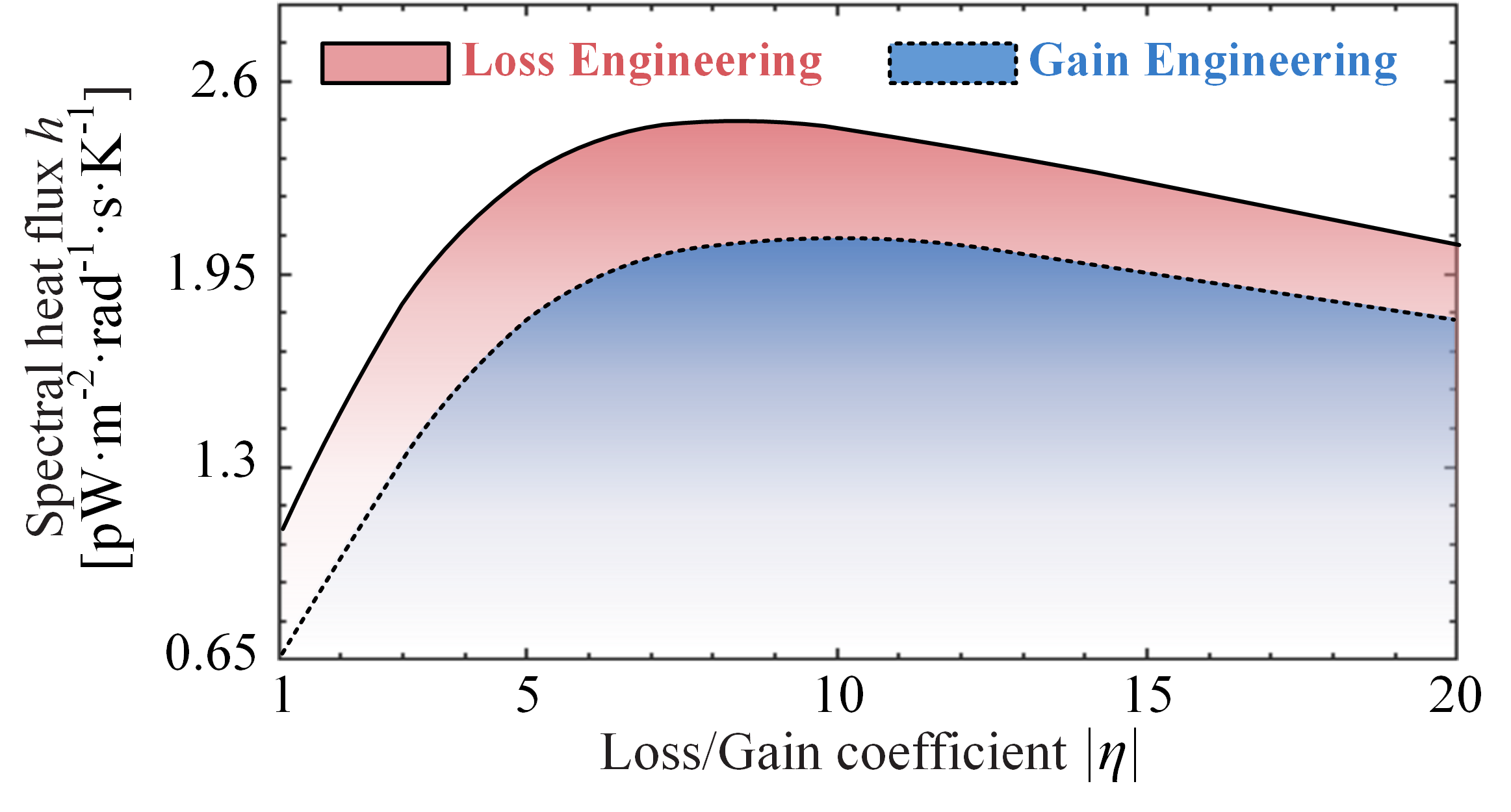}}
	\caption{Spectral heat flux $h$ as a function of gain and loss in the diagonal term of the conductivity tensor.}
	\label{Fig3}
\end{figure}

\begin{figure*}[t]
	\centering
	\centerline{\includegraphics[width=1\textwidth]{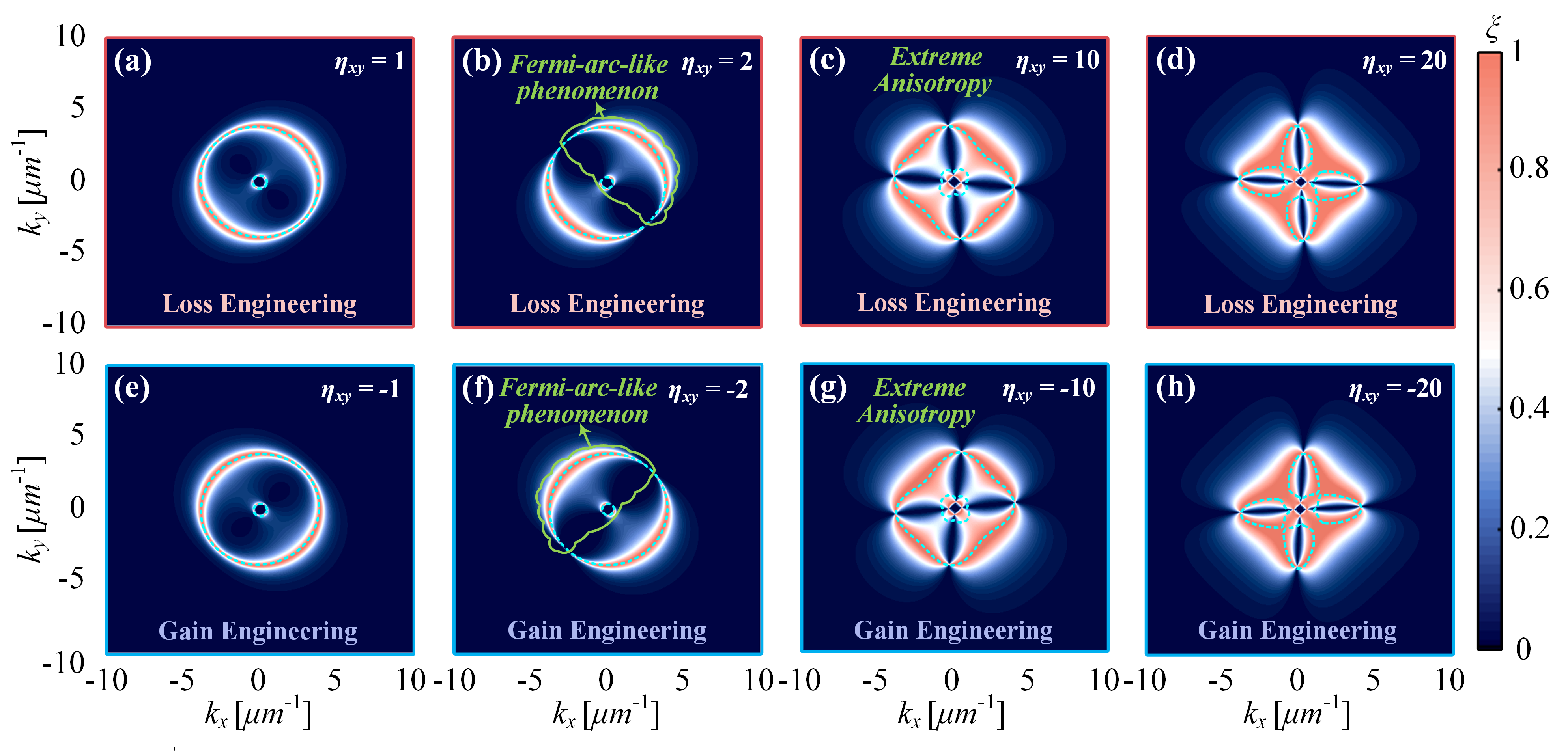}}
	\caption{Evolution of photonic transmission coefficient $\xi$ when applying gain and loss in a single non-diagonal term, parameterized by the coefficient $\eta_{xy}=\mbox{Re}(\sigma_{xy})/\mbox{Re}(\sigma_{xx})$ (with $\eta_{yx}=0$ and $\eta_{yy}=1$).
		 (a) $\eta_{xy} = 1$, (b) $\eta_{xy} = 2$, (c) $\eta_{xy} = 10$, (d) $\eta_{xy} = 20$, (e) $\eta_{xy} = -1$, (f) $\eta_{xy} = -2$, (g) $\eta_{xy} = -10$, and (h) $\eta_{xy} = -20$. The dashed curves in each panel represent the isofrequency contours.}
	\label{Fig4}
\end{figure*}

\begin{figure*}[t]
	
	\centering
	\centerline{\includegraphics[width=1\textwidth]{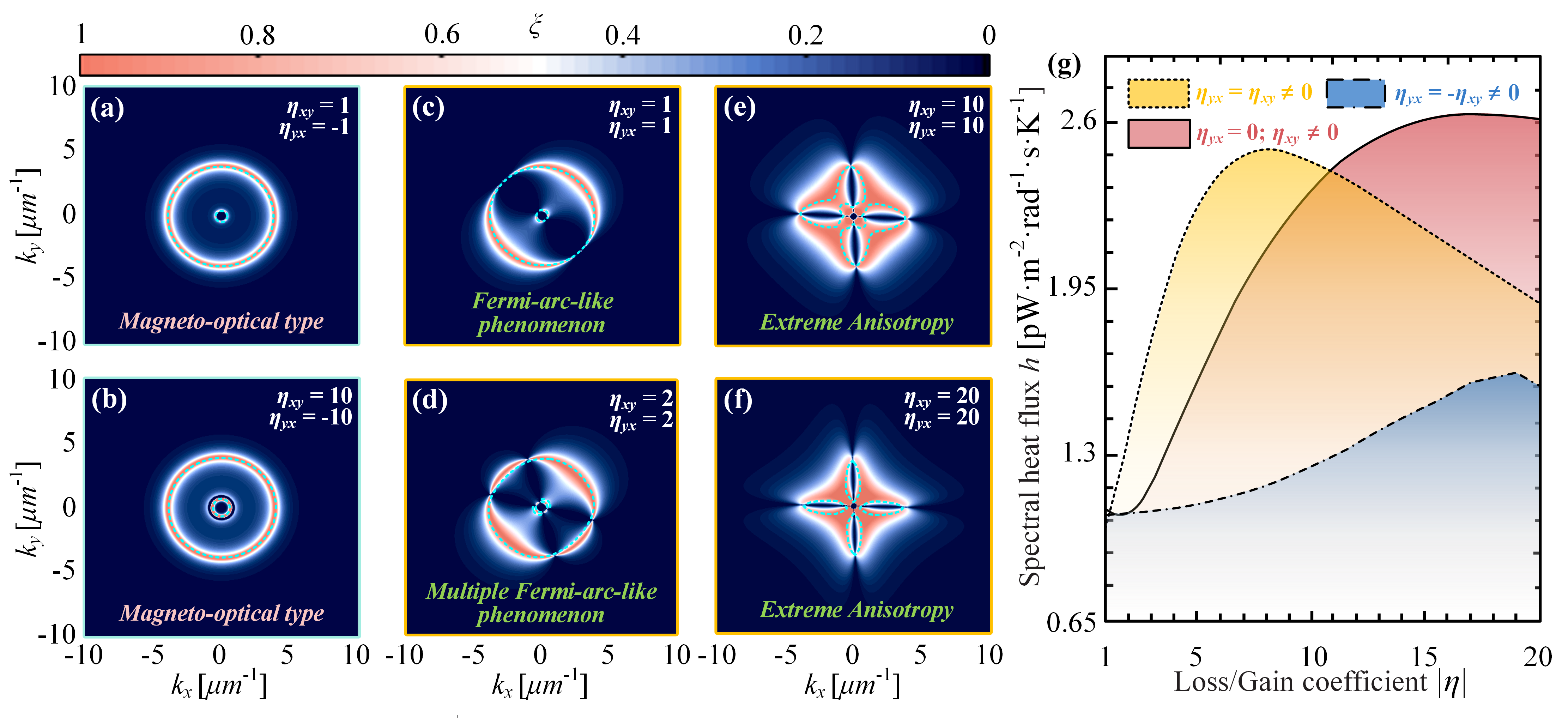}}
	\caption{Evolution of photonic transmission coefficient $\xi$ under non-diagonal gain/loss engineering. (a) $\eta_{xy} = -\eta_{yx} = 1$, (b) $\eta_{xy} = -\eta_{yx} = 10$, (c) $\eta_{xy} = \eta_{yx} = 1$, (d) $\eta_{xy} = \eta_{yx} = 2$, (e) $\eta_{xy} = \eta_{yx} = 10$, and (f) $\eta_{xy} = \eta_{yx} = 20$. The dashed curves in each panel represent the isofrequency contours. (g) Spectral heat flux $h$ as a function of non-diagonal gain-loss coefficient.}
	\label{Fig5}
\end{figure*}

We explore the thermal photon tunneling behaviors under loss and gain engineering, elucidating the origin, evolution, and practical applications of the anisotropic response, as illustrated in Figs.\,\ref{Fig2}--\ref{Fig5}. Our exploration begins with a comparison of the effects of loss-engineering [$\mbox{Re}(\sigma_{yy})>0$ and  $\neq \mbox{Re}(\sigma_{xx})$] and gain-engineering [$\mbox{Re}(\sigma_{yy})<0$ and $\neq \mbox{Re}(\sigma_{xx})$] on thermal photon tunneling, specifically focusing on the diagonal term (see Fig.\,\ref{Fig2}). Importantly, we maintain a fixed frequency of 0.05 eV and a vacuum gap of 50 nm to isolate the effects under consideration. It is worth noting that variations in frequency and vacuum gap solely affect the tunneling intensity and the range of wavevectors, without influencing the evolution of the loss-gain-induced mode. 

\setlength{\parskip}{0pt}

\setlength{\parskip}{0pt}
In our initial analysis, we consider as a reference an isotropic medium represented by a conductivity tensor with components $\sigma_{xx}=\sigma_{yy}=(0.00025 + 0.0019i$) S, $\sigma_{xy}=\sigma_{yx}=0$.
Figure \ref{Fig2}(a) displays a bright band along with its dispersion relationship (indicated by a blue dotted curve), reflecting the characteristics of an isotropic medium. Interestingly, this typical isotropic behavior of thermal photon tunneling is disrupted when loss engineering is applied, such as by increasing the real part of $\sigma_{yy}$. This modification introduces significant anisotropy, particularly constraining the tunneling in the \textit{y}-direction to a narrower range of wavevectors.

To elaborate on these effects, Fig.\,\ref{Fig2}(b) displays the in-plane dispersion for $\eta_{yy}=2$. Here, the dispersion curves exhibit elliptical anisotropy and are distinctly visible within the bright mode branches. It is also important to note that an increase in loss not only enhances the bright band of photon tunneling but may also amplify the intensity of thermal photon tunneling. The bright band is especially more pronounced along the \textit{y}-direction compared to the \textit{x}-direction, suggesting that the system's capacity to absorb thermal photons increases with loss \cite{Tsurimaki2017JQ}. This enhanced tunneling capability could prove beneficial in applications such as thermal management for high-speed electronic chips \cite{Raphael2014NL}, thermo-modulators \cite{Li2019NL}, and thermal logic circuits \cite{van2012PRL}. As the loss coefficient is further increased, as shown in Figs.\,\ref{Fig2}(c) and\,\ref{Fig2}(d), the degree of anisotropy in the system also increases.
Moreover, at high values of the loss coefficient in the diagonal term ($\eta_{yy} > 10$), we observe unique anticrossing features between the anti-symmetric (inner curve) and symmetric (outer curve) isofrequency contours \cite{ilic2012near}, adding further complexity to the thermal photon tunneling phenomena.

\setlength{\parskip}{10pt}
 When the loss coefficient increases to 20, as illustrated in Fig.\,\ref{Fig2}(d), the dispersion of thermal photon tunneling undergoes a substantial transformation, displaying quasi-hyperbolic-like characteristics. This topological shift, commonly achieved through polariton hybridization \cite{zhou2022PRM}, represents a significant advancement in the study of thermal photon dynamics. As we transition from losses to gain by changing the real part of $\sigma_{yy}$ to negative values, a new phenomenon arises. Figures \ref{Fig2}(e)--(h) demonstrate that while the introduction of gain does not alter the topological structure of the thermal photon tunneling state or its dispersion, it significantly affects the system's dynamics. Specifically, the interplay between gain and loss disrupts the previously bright band of thermal photon tunneling, resulting in the formation of a distinct thermophotonic defect state (i.e., \textit{tp}-defect state). This defect state significantly constrains the energy transport performance of the system.
 
  \setlength{\parskip}{0pt}
To quantify the thermal photon energy transfer performance at a specific frequency, we introduce the spectral heat flux
\begin{equation}
h = \frac{\partial\Theta(\omega, T)}{\partial T}\iint \xi(\omega, k_{x}, k_{y}) \,dk_{x}\,dk_{y},
\end{equation}
which serves as a measure of the system's ability to transport radiative energy at that frequency. Here, $\Theta(\omega, T) = \hbar\omega/[e^{\hbar\omega/(k_{b})T}-1]$ represents the average energy of a photon at angular frequency $\omega$, with $k_b$ denoting the Boltzmann constant \cite{zhou2022PRM}. 
As shown in Figure\,\ref{Fig3}, the introduction of both gain and loss leads to a substantial enhancement in the spectral heat flux of the system. However, counterintuitively, the introduction of gain does not result in a noticeable increase in the radiative heat transfer coefficient compared to a purely lossy system. As demonstrated in Fig.\,\ref{Fig3}, transitioning the loss/gain coefficient from $10$ (loss) to $-10$ (gain) results in a 16\% reduction in the spectral heat flux of the system. Remarkably, this anomalous decrease in radiative heat transfer deviates sharply from the usual enhancements observed in optical performance with the introduction of gain, as seen in areas such as optical transistors \cite{Rappoport2023PRL}, lasers \cite{Liu2016science}, and nonlinear optics \cite{Sirleto2012nc}.

\setlength{\parskip}{0pt}

We next investigate the complex dynamics of gain-loss-assisted metasurface systems by integrating gain and loss into the non-diagonal terms of the optical conductivity tensor. This method has been extensively used in studying the response of non-Hermitian metamaterials \cite{Passler2022n,Lannebere2022PRL}. Initially, we introduce loss into $\sigma_{xy}$ to observe the effects of non-diagonal loss-engineering [$\sigma_{xy} = \eta_{xy}\mbox{Re}(\sigma_{xx})$ and $\sigma_{yx} = 0$] on thermal photon tunneling behavior, as shown in Fig.\,\ref{Fig4}. This configuration can lead to a linear electro-optical response \cite{Rappoport2023PRL}. 
When maintaining $\mbox{Re}(\sigma_{xy})$ at smaller levels ($\eta_{xy} = 1$ and $2$), a shear effect emerges, as shown in Figs.\,\ref{Fig4}(a) and \ref{Fig4}(b). The bright band of thermal photon tunneling rotates with respect to the coordinate system, as also observed in the isofrequency contours, and undergoes significant changes in morphology. The thermal photon energy concentrates from end-to-end, resulting in a shape resembling a Fermi arc (Fermi-arc-like phenomenon) \cite{Zhou2018s}. With further increases in $\mbox{Re}(\sigma_{xy})$, the morphology of the thermal photons continues to evolve. In Fig.\,\ref{Fig4}(c), when the loss coefficient reaches $10$, the bright band of thermal photon tunneling exhibits extreme anisotropy resembling the shape of a petal. The dispersion curves become highly anisotropic and are prominently located within the bright contour of mode branches. This extreme anisotropy persists as the coefficient further increases. Additionally, in non-diagonal loss engineering, larger loss coefficients induce an anticrossing effect between symmetric and asymmetric dispersions, akin to the diagonal loss-engineering discussed earlier. As shown in Fig.\,\ref{Fig4}(d), when the loss coefficient is increased to $20$, the dispersion undergoes a topological transition, transitioning from the original two to four closed curves.

In the scenarios illustrated in Figs.\,\ref{Fig4}(e)--(h), interestingly, the inversion of gain and loss within the non-diagonal elements does not alter the fundamental mode of thermal photon tunneling; instead, it reverses its orientation. This orientation change can be explained by a coordinate transformation applied to the conductivity tensor. When the coordinates are rotated by an angle $\phi$, the transformed conductivity tensor, $\sigma_{t}$, is calculated as $\sigma_{t} = \textbf{\textit{R}}_{c}(-\phi) \sigma \textbf{\textit{R}}_{c}(\phi)$, where 
\begin{equation}
\textbf{\textit{R}}_{c}(\phi) =
\begin{bmatrix}
	\cos\phi & \sin\phi\\
	-\sin\phi & \cos\phi
\end{bmatrix}
\end{equation}
represents the rotation matrix. Switching the real part of $\sigma_{xy}$ from positive to negative effectively mimics a 90-degree rotation of the coordinate system. Consequently, as shown in Figs. \ref{Fig4}(e)--(h), modifying the real part of $\sigma_{xy}$ from positive to negative results in a 90-degree rotation in the orientation of the thermal photon tunneling mode.

Further exploration involves the simultaneous application of gain and loss to the non-diagonal terms [$\sigma_{xy} = \eta_{xy}\mbox{Re}(\sigma_{xx})$, $\sigma_{yx} = \eta_{yx}\mbox{Re}(\sigma_{xx})$, with $\eta_{xy}=-\eta_{yx}$]. This configuration effectively transforms the system into a structure resembling conventional magneto-optical metamaterials \cite{Wu2019prapp, He2019APL}. As shown in Figs.\,\ref{Fig5}(a) and\,\ref{Fig5}(b), the anisotropy of thermal photon tunneling is significantly reduced due to the interplay between non-diagonal gain and loss, in line with behaviors observed in magneto-optical systems noted in previous studies. This outcome not only corroborates the theoretical predictions but also highlights the robustness and relevance of our findings in the broader context of magneto-optical research. 
 
 \setlength{\parskip}{0pt} 
 However, when loss effects are simultaneously applied to the non-diagonal terms, anisotropy is effectively reinstated. Intriguingly, varying the real part of the non-diagonal term from positive to negative primarily affects the azimuthal angle of thermal photon tunneling, without changing its fundamental mode. Consequently, our analysis will focus on configurations where the real part of the diagonal term remains positive. In setups where $\eta_{xy}$ and $\eta_{yx}$ both equal 1, the system exhibits Fermi-arc-like phenomena, akin to effects seen with independent non-diagonal loss-engineering. As these coefficients are increased to 2, the bright-band branches of thermal photon tunneling expand, manifesting multiple Fermi-arc-like features, as observed in Fig.\,\ref{Fig5}(d). When the coefficient exceeds 10, the phenomenon of petal-like extreme anisotropy, characteristic of thermal photon tunneling, becomes pronounced, as shown in Figs.\,\ref{Fig5}(e) and\,\ref{Fig5}(f). Furthermore, the application of non-diagonal gain/loss engineering not only diversifies the anisotropic modes of thermal photon tunneling but also significantly enhances its energy transfer capabilities. Figure\,\ref{Fig5}(g) demonstrates that manipulating the gain-loss coefficients in non-diagonal terms can enhance single-frequency radiative heat transfer performance by over 200$\%$.
 
\setlength{\parskip}{20pt}
\noindent\textbf{4.\,Conclusion and outlook}

\setlength{\parskip}{5pt}
 \label{Sec4}
 In summary, our research reveals new insights into the complex interplay between gain-loss dynamics and thermophotonics, introducing a theory of anisotropic thermal photon tunneling that challenges traditional approaches in polariton engineering. This study demonstrates how the judicious modulation of gain and loss can induce a variety of anisotropic behaviors in thermal photon tunneling, including the emergence of thermal photonic defect states, as well as single and multiple Fermi-arc-like phenomena. Remarkably, our findings also uncover a counterintuitive aspect: the introduction of optical gain does not enhance energy transfer efficiency as typically seen in conventional optical devices, but rather reduces it due to the formation of a thermal photon defect state. These observations suggest a fundamental shift in the design and understanding of anisotropic thermal photon tunneling, highlighting the potential of using the full spectrum of complex conductivity values to achieve broader control and richer modal diversity in thermal photon manipulation. The superior performance in controlling thermal photon flows demonstrated here not only advances our understanding but also opens exciting avenues for practical applications in thermal management and device engineering.

 \noindent \textbf{Acknowledgements}
 
 \noindent C.-L. Z. acknowledges the support from the National Natural Science Foundation of China (Grant No. 523B2060) and the Fundamental Research Funds for the Central Universities (Grant No. HIT.DZIJ.2023098). H.-L. Y. acknowledges the support from the National Natural Science Foundation of China (Grant No. U22A20210). Y. Z. acknowledges the support from the National Natural Science Foundation of China (Grant No. 52106083). M. A. acknowledges the grant ``CAT'', No. A-HKUST604/20, from the ANR/RGC Joint Research Scheme sponsored by the French National Research Agency (ANR) and the Research Grants Council (RGC) of the Hong Kong Special Administrative Region.  V. G. acknowledges partial support from the University of Sannio through the FRA 2023 program. Part of the computational resources were provided by the DIPC computing center.
 
\noindent\textbf{Conflict of Interest}

\noindent The authors declare no conflict of interest.

\noindent\textbf{Data Availability Statement}

\noindent The data that support the findings of this study are available from the corresponding authors upon reasonable request.

\noindent \textbf{Keywords}

\noindent Thermal phonon tunneling, near-field radiative heat transfer, extreme anisotropy, metasurfaces. 

\bibliography{reference}

\end{document}